\documentclass[12pt]{JHEP3}

\usepackage{epsfig}

\usepackage{amsmath}
\usepackage{amssymb}
\usepackage{cite}
\newcommand{\beqs}{\begin{equation*}}
\newcommand{\beq}{\begin{equation}}

\newcommand{\eeqs}{\end{equation*}}
\newcommand{\eeq}{\end{equation}}

\newcommand{\beqas}{\begin{eqnarray*}}
\newcommand{\beqa}{\begin{eqnarray}}

\newcommand{\eeqas}{\end{eqnarray*}}
\newcommand{\eeqa}{\end{eqnarray}}

\newcommand{\eq}[2]{\begin{equation} #1 \label{#2} \end{equation}}

\newcommand{\meq}[2]{\begin{multline} #1 \label{#2} \end{multline}}

\newcommand{\eps}{\varepsilon}

\newcommand{\de}{\delta}
\newcommand{\om}{\omega}
\newcommand{\ka}{\kappa}
\newcommand{\la}{\lambda}
\newcommand{\si}{\sigma}

\newcommand{\Ga}{\Gamma}
\newcommand{\De}{\Delta}

\newcommand{\blist}{\begin{itemize}}

\newcommand{\elist}{\end{itemize}}

\providecommand{\href}[2]{#2}

\newcommand{\twod}{$2D$}        % with/without $, d or D - I don't care; please use {\twod} in the text!
\newcommand{\spec}{C_s} 

\newcommand{\nilcv}{vanishing inverse specific heat}

\title{Positive specific heat of the quantum corrected dilaton black hole}

\author{D.\ Grumiller, W.\ Kummer\\ Institut f\"ur Theoretische Physik, TU Wien, \\ Wiedner Hauptstr.\  8-10, A-1040 Wien, Austria\\ E-mail: \email{grumil@hep.itp.tuwien.ac.at}, \email{wkummer@tph.tuwien.ac.at}}

\author{D.V.\ Vassilevich\thanks{On leave from V.Fock Institute of Physics, St.Petersburg University, 198904 St.Petersburg, Russia}\\ Institut f\"{u}r Theoretische Physik, Universit\"{a}t Leipzig,\\ Augustusplatz 10, D-04109 Leipzig, Germany \\ E-mail: \email{vassil@itp.uni-leipzig.de}}

\abstract{Path integral quantization of dilaton gravity in two dimensions is applied to the CGHS model to the first nontrivial order in matter loops. Our approach is background independent as geometry is integrated out exactly. The result is an effective shift of the Killing norm: the apparent horizon becomes smaller. The Hawking temperature which is constant to leading order receives a quantum correction. As a consequence, the specific heat becomes positive and proportional to the square of the black hole mass.}


\preprint{TU--03--12}

\begin{document}

\section{Introduction}\label{se:1}

The past three decades have seen remarkable progress towards a quantum theory of gravity\footnote{For a recent survey with focus on loop quantum gravity and string theory cf.\ \cite{Smolin:2003rk}.}. Many results have been obtained first in simpler twodimensional (\twod) models, because although one encounters essentially the same conceptual problems the technical ones are less demanding. Especially in the context of classical and quantum physics of black holes (BHs) {\twod} models provide a very useful laboratory to study basic questions (cf.\ \cite{Grumiller:2002nm} for a recent review).

Among these models the one of Callan, Giddings, Harvey and Strominger (CGHS) \cite{Callan:1992rs} has been (and continues to be) particularly popular. It exhibits a BH solution, called the dilaton BH, which originally arose in the context of string theory \cite{Mandal:1991tz,Dijkgraaf:1992ba}. We mention just two of its peculiar features: it is soluble exactly at the classical level, despite of the presence of minimally coupled matter, and the Hawking temperature is independent of the BH mass. One should think that in view of the huge amount of literature \cite{Giddings:1994pj,deAlwis:1992as,Bose:1995pz,Russo:1992ht,Mikovic:1992id,Mikovic:1997de,Kuchar:1997zm} by now essentially everything is known about it, but this is not the case because classical solubility does not imply quantum solubility. 

In this paper we will pose (and answer) a simple physical question: what is the specific heat of the dilaton BH?

The answer derived from the standard Hawking temperature law yields a trivial result (mass and temperature are independent of each other); thus, the specific heat encodes effects which go beyond the straightforward applications of quantum field theory on a fixed curved background. Fortunately, in {\twod} gravity with matter a scheme exists \cite{Haider:1994cw,Kummer:1997hy,Kummer:1998zs,Grumiller:2000ah,Grumiller:2001ea} in which geometry is taken into account nonperturbatively. 

The philosophy of that approach is to integrate out geometry first {\em exactly} (which is always possible in {\twod}) and then to apply a perturbative expansion in the ensuing effective theory which depends solely on matter degrees of freedom and external sources. To each given loop order and order in the matter fields geometry can be reconstructed unambiguously.

We will prove that an application of this background independent quantization scheme to the first nontrivial order in matter for the specific heat of a dilaton BH with mass $M$ yields
\eq{
\spec := \frac{dM}{dT_H} = \frac{96\pi^2}{\la^2}M^2\,,
}{cghs:1}
which shows the same mass dependence but the opposite sign as compared to the specific heat of the Schwarzschild BH. 

A remark on the matter loop counting is in order. There are two classes of approaches
\cite{Frolov:1998} to derive the Hawking temperature. The first 
one considers the tree-order scattering on the classical BH background
and relates the Hawking temperature to the Bogoliubov coefficients between ingoing and outgoing states. In the present work we calculate one-loop corrections to this classical background. The second approach is based upon the study of vacuum expectation values of the energy-momentum tensor. The leading order in the latter is produced by the one-loop matter contributions. In this sense, we calculate two-loop corrections. In any case we obtain a correction to the otherwise trivial specific heat.

This paper is organized as follows: in section 2 the path integral quantization of the CGHS is reviewed briefly with focus on the main nontrivial steps. Section 3 contains the new results which eventually lead to the expression
(\ref{cghs:1}) for the specific heat. The discussion and possible 
generalizations are contained in section 4. Complementary material 
can be found in the appendix.

\section{Path integral quantization of CGHS}

The purpose of this section is to sketch the main steps \cite{Haider:1994cw,Kummer:1997hy,Kummer:1998zs,Grumiller:2000ah,Grumiller:2001ea} of the path integral quantization within the first order formulation of dilaton gravity. 

The point of departure is the CGHS action, which in our notation\footnote{Instead of considering $N$ scalars like in the original work we just take a single field; adding more scalars does not change anything essential in our approach.
The action (\ref{cghs:2}) differs by an overall factor of $1/\pi$ from ref.\ \cite{Callan:1992rs} and by a relative factor of $-2$ between geometric and matter part. We use the sign conventions and normalization factors of \cite{Grumiller:2002nm}, except for the overall sign of the Lagrangian.\label{fn:1}} reads
\eq{
L=\frac{1}{2}\int d^2x\sqrt{-g}\left[XR+\frac{(\nabla X)^2}{X}-4\la^2X+(\nabla\phi)^2\right]\,,
}{cghs:2}
where $g,R,X,\phi$ are (determinant of the) metric, curvature scalar, dilaton and scalar field, respectively. The scale parameter $\la^2$ just fixes the physical units. For definiteness it will be assumed that $\la\in\mathbb{R}^+$.

The first crucial step is to realize that (\ref{cghs:2}) is classically \cite{Katanaev:1996bh} and quantum mechanically \cite{Kummer:1997hy} equivalent to a first order formulation in terms of Cartan variables  and auxiliary fields:
\eq{
L^{(1)}=-\int \left[X_a (D\wedge e)^a +Xd\wedge\omega -\epsilon(2\la^2X+X^+X^-/X)+\frac12 d\phi\wedge\ast d\phi) \right]
}{cghs:3}
$e^a$ is the zweibein one-form,
$\epsilon$ is the volume two-form. The one-form $\omega$ represents the 
spin-connection $\om^a{}_b=\eps^a{}_b\om$ with $\eps_{ab}$ being the totally
antisymmetric Levi-Civit{\'a} symbol. The kinetic term of the scalar field $\phi$ contains the Hodge $\ast$ operation. The action (\ref{cghs:3}) 
depends on two auxiliary fields $X^a$. Its geometric part is a special case of a 
Poisson-$\si$ model \cite{Ikeda:1993aj}
with a three dimensional target space the
coordinates of which are $X,X^a$. With flat metric $\eta_{ab}$ in light-cone coordinates ($\eta_{+-}=1=\eta_{-+}$, $\eta_{++}=0=\eta_{--}$) the first (``torsion'') 
term of (\ref{cghs:3}) is given by
\begin{equation}
X_a(D\wedge e)^a = \eta_{ab}X^b(D\wedge e)^a =X^+(d-\omega)\wedge e^- +
X^-(d+\omega)\wedge e^+\,.\label{dvvXDe}
\end{equation}
Below a certain (non-covariant) gauge will be employed, so we have to coordinatize the 1-forms $e^a=e^a_\mu dx^\mu=e^a_0dx^0+e^a_1dx^1$. Partial derivatives always refer to these coordinates, e.g.\ $\partial_0:=\partial/\partial x^0$. Despite the suggestive notation $x^0$ need not be identified necessarily with a ``time'' variable. Indeed, below $x^0$ will rather play the role of a radial coordinate explaining the somewhat unusual overall minus sign in (\ref{cghs:3}).
%\footnote{This is the reason why (\ref{cghs:14}) acquires a somewhat unusual overall sign. See also footnote \ref{fn:1}.}

This reformulation is already very convenient classically \cite{Klosch:1996fi}, but it becomes crucial at the quantum level. The path integral is obtained by means of standard BRST methods \cite{Fradkin:1975cq}
\begin{align}
W &= \int\, (\mathcal{D} \om_\mu )\,(\mathcal{D} e^a_\mu )\,   
(\mathcal{D} X ) \, (\mathcal{D} X^a )\, (\mathcal{D} \tilde{\phi} )\, 
(\mathcal{D} {\rm ghost} )\, \nonumber \\
& \quad \times \exp\; \left[ 
i  (L^{(1)} +  L^{({\rm g})} + \rm sources )\,\right]\, ,
\label{eq:4.29}
\end{align}
with $L^{(1)}$ given by (\ref{cghs:3}) and $L^{(\rm g)}$ containing the gauge fixing part and the ghost-sector; all corresponding path integrations have been formally lumped together into the term $\mathcal{D} {\rm ghost}$. The correct measure for the scalar field $\mathcal{D}\tilde{\phi}$ and the source-terms will be specified below. 

It is convenient to employ the gauge 
\eq{
e_0^+=0=\om, \quad e_0^-=1\,, 
}{dghs:1}
which eventually will lead to a line element of Sachs-Bondi type (cf.\ eq.\ (\ref{line}) below). Having integrated out the ghost and gauge-fixing sector one obtains \cite{Kummer:1997hy,Kummer:1998zs,Grumiller:2001ea}
\begin{equation}
W = \int\, (\mathcal{D} \om_1)(\mathcal{D} e^a_1)(\mathcal{D}X)(\mathcal{D}X^a)(\mathcal{D}(\phi {e_1^+}^{1/2})) \det \De \, \exp\, i L^{(2)}\,,
\label{eq:4.35}
\end{equation}
where the measure contains now a determinant $\det \De$ depending solely on the target space variables. The quantity $\mathcal{D}\tilde{\phi}$ has been fixed such that the proper diffeomorphism covariant measure $\mathcal{D}(\phi(-g)^{1/4})=\mathcal{D}(\phi {e_1^+}^{1/2})$ is reproduced \cite{Fujikawa:1988ie}. The action in (\ref{eq:4.35}) is given by
\meq{
L^{(2)} = \int d^2x \Big[ -X\partial_0\om_1-X_a\partial_0e^a-\om_1 X^+-e_1^+(2\la^2X+X^+X^-/X) \\
+\left((\partial_1\phi)(\partial_0\phi)-e_1^-(\partial_0\phi)^2\right)+j_1\om_1+j_2e^-_1+j_3e^+_1+J_1X+J_2X^++J_3X^-+\sigma\phi\Big]\,.
}{eq:4.36}
The sources $\si,j_i,J_i$ for the scalar field, the remaining Cartan variables and the target space coordinates now have been specified. It is important to realize that (\ref{eq:4.36}) is linear in the Cartan variables. However, the appearance of ${e_1^+}^{1/2}$ in the measure prohibits an immediate path integration. This problem is solved by introducing a new auxiliary field $f$ and exploiting the fact that $j_3$ is the source of $e_1^+$. Then (\ref{eq:4.35}) can be represented as
\eq{
W=\int(\mathcal{D}f)\delta\left(f-\frac1i \frac{\delta}{\delta j_3}\right) \widetilde W \,,
}{cghs:4}
where in $\widetilde W$ the factor $\mathcal{D}(\phi{e_1^+}^{1/2})$ has been replaced by $\mathcal{D}(\phi f^{1/2})$, but otherwise it is identical to (\ref{eq:4.35}). 

Thus, path-integrations over the Cartan variables just yield $\de$-functions, which can be used to perform the path-integrations over the target space coordinates. Incidentally, the functional determinant obtained in this way just cancels\footnote{This cancellation might be anticipated on general grounds and has been checked later in the framework of reduced phase space quantization \cite{Katanaev:2000kc}.} $\det \De$. However, one has to be careful with the evaluation of the $\de$-functions as their arguments yield three linear first order differential equations and thus homogeneous contributions depending on $x^1$ do exist. We drop the sources $j_1,j_2$ from now on as they are inessential for the perturbation theory in the scalar field discussed below. The solutions of these differential equations for the target space coordinates can be written as
\begin{gather}
X = X_{\rm hom} + \partial_0^{-1} X^+ \, , \label{eq:4.39}\\
X^+ = X^+_{\rm hom} + \partial_0^{-1} (\partial_0 \phi)^2 \, , \label{eq:4.40}\\
X^- = X\left[\partial_0^{-1}\frac{1}{X}(-j_3+2\la^2X)-M\right] \,, 
\label{eq:4.41}
\end{gather}
with the homogeneous solutions $\partial_0X_{\rm hom}=\partial_0X^+_{\rm hom}=\partial_0M=0$. As we will see later $M$ is nothing but the BH mass. Of course, the operators $\partial_0^{-1}$ still have to be defined properly. This important issue will be postponed until we actually start calculating Green functions; however, one particular consequence of their appearance, namely an ambiguity in the action, will be addressed immediately because it is rather important: the terms in (\ref{eq:4.36}) proportional to $J_i$ contain contributions with $\partial_0^{-1}$ when the solutions (\ref{eq:4.39}-\ref{eq:4.41}) are inserted. In an expression $\int dx^0 \, \int dy^0 \, J_{x^0} (\partial_0^{-1})_{x^0y^0} A_{y^0}$ the  symbol $\partial_0^{-1}$ means an integral which when acting upon $J$ contains an undetermined integration constant $\bar{g} (x^1)$. This generates a new term $\bar{g}\, \int A$. Similarly, for $\partial_0^{-2}$ a term linear in $x^0$ is produced. The terms $J_1X+J_2X^+$ in this way produce an additional contribution to the action (\ref{eq:4.36}) which reads (modulo field independent terms)
\eq{
(g_1+g_2x^0)(\partial_0\phi)^2\,.
}{cghs:5} 
The functions $g_1(x^1)$ and $g_2(x^1)$ are fixed by requirements on the asymptotic behavior of the effective line element: neglecting any backreaction or loop effects in the gauge (\ref{dghs:1}) this term has to produce the classical kinetic term $-E_1^-(\partial_0 \phi)^2$ with
\eq{
E_1^-=-\frac{M}{2\la}+\la x^0\,,
}{killing}
where $M$ can be identified with a constant BH mass.\footnote{Cf.\ the first two terms in (3.15) of ref.\ \cite{Grumiller:2002nm} with $M=-m_\infty$ and $w(x^0)=-2\la^2x^0$ for the CGHS model. There is, however, a subtlety concerning the mass dimensions: $X^\pm, X$ have mass dimension one and zero, respectively. Thus, the quantity $X^+_{\rm hom}$ also has mass dimension of one. However, in previous publications (cf.\ \cite{Grumiller:2002dm} and some of the references therein) it has been fixed to $X^+_{\rm hom}=1$ because natural units have been employed, i.e.\ quantities like $\la$ in the present paper have been fixed to a dimensionless constant. In the present case we require $X^+_{\rm hom}=2\la$, because the conversion factor between the Casmir function and the BH mass for the CGHS model is given by eq.\ (5.13) of \cite{Grumiller:2002nm}. In the absence of matter this implies  $X=2\la x^0$.} Actually (\ref{cghs:5}) would permit also a dependence $M(x^1)$. We will return to this point at the end of section \ref{se:3}. The symbol $E_1^-$ indicates that this term replaces $-e_1^-(\partial_0\phi)^2$ which appeared in (\ref{eq:4.36}) before $e_1^-$ has been integrated out.

The term $J_3X^-$ in general establishes the most important contribution 
\eq{
g_3\frac{1}{X}\left(j_3-2\la^2X\right)\,,
}{cghs:6}
because it implies nontrivial interactions of the scalar with geometry. For the CGHS model the source independent second term in (\ref{cghs:6}) is irrelevant as opposed to the situation in generic dilaton gravity models.\footnote{In the generic case these source independent terms are responsible for classical vertices with $2n$ outer legs and they can be used to describe scattering processes \cite{Grumiller:2000ah,Fischer:2001vz,Grumiller:2002dm}. This is also the reason why the terms appearing in (\ref{cghs:5}) did not have to be considered separately in ref.\ \cite{Kummer:1998zs}. They are only relevant for the very special class of dilaton gravity models without classical vertices to which the CGHS belongs.} The function $g_3(x^1)$ essentially becomes a scale factor in front of $e_1^+$ (because the classical value of $e_1^+$ is just $1/X$ and $j_3$ is its source), so it can be fixed to unity. From now on the sources $J_i$ can be set to zero. In addition to the matter source $\si$ only $j_3$ (which is essentially the source for the determinant of the metric) has to be kept because of (\ref{cghs:4}).
 
After performing these steps the generating functional for Green functions can be presented as
\begin{equation}
W[\si]=\left.\int (\mathcal{D}f) \delta\left(f-
\frac1i \frac{\delta}{\delta j_3}\right) \widetilde W [f,j_3,\si]\right|_{j_3=0}
\label{tilS}
\end{equation}
with the functional $\widetilde W$ being given as 
\begin{eqnarray}
\label{Wtilde}
\widetilde{W} [f,j_3,\si]&=&\int (\mathcal{D}\widetilde \phi) \exp i L^{\mbox{\scriptsize{eff}}} \\
\label{Wtilde-b}
L^{\mbox{\scriptsize{eff}}} &=&  \int d^2x \left[(\partial_0 \phi)(\partial_1 \phi) - E_1^-(\partial_0 \phi)^2 + j_3 \hat{E}_1^+ + \phi\si\right] 
\end{eqnarray}
where $E_1^-$ is determined by (\ref{killing}). The integration variable in (\ref{Wtilde}) is $\widetilde \phi=f^{1/2} \phi$. This yields at 1-loop level an effective action of the Polyakov type \cite{Polyakov:1981rd} to be dealt with in detail in the next section. Formally $\hat{E}_1^+$ reads\footnote{Our suggestive notation already indicates that the effective zweibein $E_1^+:=\hat{E}_1^+(\phi=0)$ receives corrections from backreactions by the scalar field.}
\eq{
\hat{E}_1^+=\left(2\la x^0+\partial_0^{-2}(\partial_0\phi)^2\right)^{-1}\,.
}{B12}
At this point all backreactions are still taken into account in a self-consistent manner. For later applications the notions
\eq{
E_1^+:=\frac{1}{2\la x^0}\,,\quad (E_1^+)^{(1)}:=-\frac{1}{(2\la x^0)^2}\partial_0^{-2}(\partial_0\phi)^2\,,
}{cghs:8}
turn out to be helpful. They are the first two terms in a powers series of $\hat{E}_1^+$ in terms of $(\partial_0\phi)^2$.

In our approach to quantization the presence of a BH followed from the path integral {\em without} the introduction of a background. However, the boundary conditions on the auxiliary fields -- already included in (\ref{tilS}-\ref{B12}) -- induce a similar effect: they determine the BH mass $M$ and the behavior of $\hat{E}_1^+$ in the absence of matter; other choices of $M, X_{\rm hom}, X^+_{\rm hom}$ would correspond to different BH masses, rescalings and shifts of the coordinate $x^0$, respectively. Indeed, all boundary values and ambiguities have been fixed uniquely by referring to the asymptotic region in the classical limit. In the sum over histories only such paths are considered which are consistent with the boundary conditions imposed on the auxiliary fields. In physical terms an asymptotic observer is assumed to live in a fixed topological sector and to measures a fixed ADM mass; this explains why no sum over topologies occurs in (\ref{eq:4.29}). 

Reconstructing the classical geometry, the effective line element $(ds)^2=2E_1^+ dx^1 \cdot (dx^0 + E_1^-dx^1)$, after a coordinate redefinition $u=x^1$, $r=\ln{(2\la x^0)}/(2\la)$, leads to
\eq{
(ds)^2=2drdu+\xi(r)(du)^2\,,\quad\xi(r):=\left(1-\frac{M}{\la}e^{-2\la r}\right)\,,
}{line}
with Killing norm $\xi(r)$. The asymptotic region is located at $r\to+\infty$, resp.\ $x^0\to+\infty$ and the singularity is encountered at $r\to-\infty$, resp.\ $x^0\to 0$.

\section{Matter loop effects}\label{se:3}

The term $j_3\hat E_1^+$ in (\ref{Wtilde-b}) provides the only nontrivial interaction of the scalars with geometry in the CGHS model. The basic idea is to perform the path integration for the quadratic part of $\phi$ in (\ref{Wtilde}) and to consider higher powers in $\phi$ perturbatively. Without interaction and without source term the integral is of the type
\eq{
\int(\mathcal{D}\phi f^{1/2})\exp{i\int d^2xf\left[\frac{1}{2}g^{\mu\nu}(\partial_\mu\phi)(\partial_\nu\phi)\right]} = e^{iL_P}\,,
}{cghs:13}
where Polyakov's effective action is given by \cite{Polyakov:1981rd}
\eq{
L_P=\frac{1}{96\pi}\int_x\int_y f R_x \square^{-1}_{xy}R_y
}{cghs:14}
with $\square:=g^{\mu\nu}\nabla_{\mu}\partial_\nu$ and $R$ being the curvature scalar of the background geometry $g_{\mu\nu}$. In the gauge (\ref{dghs:1}) the components of the inverse metric are given by
\eq{
g^{01}=f^{-1}=g^{10}\,,\quad g^{00}=-2f^{-1}E_1^-\,,\quad g^{11}=0\,,
}{cghs:15}
with $E_1^-$ as defined in (\ref{killing}) and the wave operator becomes
\eq{
\Ga:= \frac{1}{2}f\square = \left(\partial_1-\partial_0(\la x^0-\frac{M}{2\la})\right)\partial_0\,.
}{1loop8}
Now the 1-loop effective action (\ref{cghs:14}) in the gauge (\ref{dghs:1}),
\eq{
L_P=\frac{1}{48\pi}\int_x\int_y (\partial_0^2E_1^--\Ga\ln{f})_x\Ga_{xy}^{-1}(\partial_0^2E_1^--\Ga\ln{f})_y\,,
}{cghs:16} 
for the CGHS reduces to a local quantity\footnote{In the sense that no propagator terms $\Ga^{-1}$ are present.} because of (\ref{killing})
\eq{
L_P=\frac{1}{48\pi}\int \ln f \left(\Ga \ln f\right)\,. 
}{1loop7}
Performing the integration $\mathcal{D}f$ to leading order just implies the replacement $f\to E_1^+$ in (\ref{cghs:13})-(\ref{1loop7}), i.e.\ the result of quantization on a fixed background. Clearly our intention is to proceed beyond that level. 

There are three steps to be performed: A) integrate out the scalars thus obtaining the 1-loop effective action; B) integrate out the auxiliary field, thus replacing $f$ by $\hat{E}_1^+$; C) expand $\hat{E}_1^+$ in powers of $\phi$ (or in powers of $-i\de/\de\si$). The  order is essential since, for instance, integration over the scalars can hardly be performed after integration over the auxiliary field because the measure would contain nonpolynomial factors in $\phi$. Thus A) must be performed before B). To perform A) before C) also is advantageous in order to impose perturbation theory at the last possible instant. 

Reinserting the interaction term $j_3 \hat{E}_1^+$ and the source 
term $\si\phi$ and performing the path-integration over $\tilde{\phi}$ 
for the generating functional (\ref{Wtilde}) yields
\eq{
\widetilde{W} [f,j_3,\si] = \exp \left[i\int j_3\hat{E}_1^+\left(\phi\rightarrow\frac{\de}{i\de\si}\right)\right] \cdot \exp \left[i \left(L_P+\frac{1}{4}\int\si\Ga^{-1}\si\right)\right]\,. 
}{cghs:7}
The focus will be on the next to leading order term which effectively produces an interaction of a pair of scalars with the Polyakov loop, because the $\mathcal{D}f$ integration essentially replaces $f$ in (\ref{1loop7}) by $\hat{E}_1^+(\phi\rightarrow-i\de/\de\si)$ (expanded in powers of $\phi$ up to the investigated order, i.e.\ $\phi^2$).

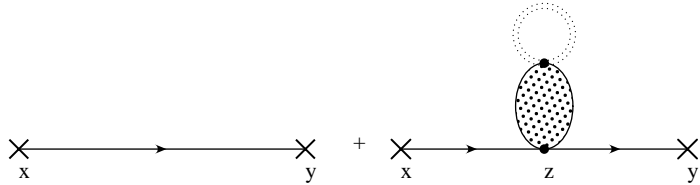
\EPSFIGURE{cghs1loop.epsi}{Propagator plus correction term}
In fig.\ 1 the relevant Feynman diagrams are depicted. The first term contains just the propagator $\Ga^{-1}$ related to (\ref{1loop8}), while the second one encodes the Polyakov-loop induced correction (the double line marks the part of the scalars which has been integrated out) between two such propagators. The formula corresponding to these diagrams is the usual one, $-\frac1W\frac{\partial W}{\partial\si_x\partial\si_y}|_{\si=0}=-\frac{\partial}{\partial\si_x\partial\si_y}[(1+iV+\dots)(1+\frac i4\int\si\Ga^{-1}\si-\frac{1}{32}(\int\si\Ga^{-1}\si)(\int\si\Ga^{-1}\si)+\dots)]|_{\si=0}$, where $V$ is the vertex calculated below (cf.\ eqs.\ (\ref{cghs:17}) and (\ref{cghs:18})). It contains two derivatives with respect to the source $\si$. Moreover, it contains two $\partial_0$ derivatives .
The interpretation is as follows: a scalar field interacts via the Polyakov loop with itself. Since the external legs are actually of the form $(\partial_0\phi)$ rather than just $\phi$ it is to be expected that the Killing norm effectively acquires a nontrivial correction which is determined to a large extent by the renormalization prescription implicit in the Polyakov action (\ref{1loop7}). 

In the appendix a more straightforward but also more lengthy derivation for the simpler case of $M=0$ is presented. That method is suitable for generic dilaton gravity. However, the CGHS allows for substantial simplifications already from the very beginning.
 
By virtue of (\ref{cghs:8}) the interaction term to order $\phi^2$ is
\eq{
V=-\frac{1}{24\pi}\int_x\int_y \left(\frac{1}{2\la x^0}\Ga\ln{E_1^+}\right)_x(\partial_0^{-2})_{xy}(\partial_0\phi)_y^2\,,
}{cghs:17} 
where strictly speaking $\phi$ should be replaced by $-i\de/\de\si$. Naively, one might expect a symmetric expression of the type (first order) $\times(\Ga$ zeroth order)$ + $(zeroth order)$ \times(\Ga$ first order), rather than just twice (first order) $\times(\Ga$ zeroth order) as in (\ref{cghs:17}). However, this is another instance where special care must be taken when an argument is based upon an effective action of Polyakov type the reliable basis of which being the conformal anomaly. More precisely, we use that to first order the Polyakov action should read $\int (\delta E_1^+)(\delta L_P/\de E_1^+)$, where $(\delta L_P/\de E_1^+)$ is given by the conformal anomaly. This property actually {\it defines} the Polyakov action and must be considered as more fundamental than its non-local expression (\ref{cghs:14}) (cf. \cite{Dowker:1994rt} for a discussion on this point). Indeed, the symmetric form would not produce the correct conformal anomaly $\Ga\ln{E_1^+}$ and as a consequence the vacuum expectation value of the propagator would receive a nontrivial correction. This feature is demonstrated more explicitly in the appendix.

There are several ways to deal with the integral kernel in (\ref{cghs:17}). The simplest seems to be to act with the double integral on the $x^0$-dependent term and to fix the integration constants to zero on the basis that no change of the asymptotics should occur. This yields immediately
\eq{
V=-\frac{1}{192\pi} \int_y \frac{M}{\la^2 y^0}(\partial_0\phi)^2_y\,.
}{cghs:18}
Thus the interaction of a pair of scalar fields with the Polyakov loop produces a (local) vertex which is proportional to the BH mass and couples only to the $(\partial_0\phi)^2$ part of the kinetic term.

But this means effectively a shift of $E_1^-\to E_1^-+M/(192\pi\la^2 x^0)$ in (\ref{killing}). Together with (\ref{cghs:8}) and a coordinate redefinition as in (\ref{line}) this modifies the classical Killing norm in the line element (\ref{line}) to
\eq{
\hat{\xi}=1-\frac{M}{\la}e^{-2\la r}+\frac{M}{48\pi\la}e^{-4\la r}\,,
}{1loop101}
and (to this order) matter fields just propagate on this shifted ``background'' geometry. This somewhat resembles the situation in semiclassical approaches where the quantum effects provide corrections to an otherwise classical calculation.\footnote{It is possible to generate the correction term in (\ref{1loop101}) by a redefinition of the dilaton dependent potentials in (\ref{cghs:2}). A simple calculation shows that, for instance, the potential in front of the kinetic term for the dilaton is shifted as follows: $1/X\to 1/X-1/(48\pi X^2)$.} That something similar happens here too -- despite of our background independent quantization -- is due to the fact that in the CGHS model only a local self interaction is produced (cf.\ fig.\ 1). The correction term becomes dominant only close to the singularity where our approximation breaks down. The Killing norm (\ref{1loop101}) exhibits now two horizons: one close to the classical one ($r=r_h$) and the other close to the singularity; however, the latter should be only an artifact of our approximation -- it anyhow has no effect upon the asymptotic Hawking flux.

Now well-known methods can be applied to extract the corrected Hawking temperature\footnote{The simplest possible definition is through the surface gravity $\ka:=\frac12 d\hat\xi/dr|_{r=r_h}=2\pi T_H$; cf.\ e.g.\ ref.\ \cite{Kummer:1999zy} and references therein.}
\eq{
T_H=T_H^0\left(1-\frac{\la}{48\pi M}\right)\,,
}{1loop102}
where $T_H^0=\la/(2\pi)$ is the ``classical'' Hawking radiation. This formula holds as long as $\la \ll M$, i.e.\ the 1-loop approximation is valid. The specific heat is then given by (\ref{cghs:1}) which concludes the proof of our main statement. 

It should be noted that in principle one could use also the background field formalism to calculate perturbative corrections to the static BH, starting from the nonlocal zweibein (\ref{cghs:8}). However, our method seems to be simpler as it allows to avoid complicated problems of treating the nonlocalities of the Polyakov action. The relation of the present approach to the background field formalism has been discussed also in ref.\ \cite{Kummer:1998zs}.

Finally, we would like to address the issue of $M$ in (\ref{killing}). Like the other ``constants'' in (\ref{eq:4.39})-(\ref{eq:4.41}) it is related to a residual gauge freedom left by the condition (\ref{dghs:1}). So far it has been fixed to a constant, but in principle it could be any function of $x^1=u$ as well. In the absence of matter it essentially coincides with the ADM mass (whenever this notion makes sense -- for the CGHS model it does) and it has to be constant even at quantum level \cite{Kummer:1998zs}. In the presence of matter it can be identified with the (outgoing) Bondi mass \cite{Grumiller:1999rz} and thus becomes in general a function monotonically decreasing with $u$. Within the present approach to leading order the Bondi mass coincides with the ADM mass because of the imposed large BH approximation $\la\ll M$. The same condition ensures the validity of the 1-loop approximation. Since the leading order Hawking temperature is mass independent any correction coming from a varying Bondi mass will not be relevant until next-to-next-to leading order. 

Nevertheless, it is amusing that all steps leading to the result (\ref{1loop102}) remain valid even for non-constant $M=M(u)$. Thus, one can calculate the leading order mass decrease by virtue of the $2D$ Stefan-Boltzmann law:
\eq{
\frac{dM}{du}= -\frac{\pi}{6} (T_H^0)^2\left(1-\frac{\la}{24\pi M}\right) + {\cal O}\left(\frac{\la^2}{M^2}\right)
}{referee:1}
Supposing that at $u=u_0$ the initial mass of the BH was $M=M_0$ this equation can be integrated straightforwardly
\eq{
M(u)\approx M_0-\frac{\pi}{6} (T_H^0)^2(u-u_0)+\frac{\la}{24\pi}\ln{\frac{M(u)-\la/(24\pi)}{M_0-\la/(24\pi)}}\,.
}{referee:2}
The first term is the ADM mass, the second term corresponds to a linear decrease due to the (in leading order) constant Hawking flux and the third term provides the first nontrivial correction. Up to the considered order in $\la/M$ from (\ref{referee:2}) one can express $M(u)$ in terms of the Lambert $W$-function \cite{Corless:1996}. 

\section{Discussion}

Inspection of the Killing norm (\ref{1loop101}) reveals that the horizon is shifted to a slightly smaller value of the dilaton due to backreaction effects. Consequently, the Hawking temperature (\ref{1loop102}) decreases as compared to leading order, which may be an indication that eventually the Hawking process will stop for the dilaton BH. Of course the zero present in (\ref{1loop102}) cannot be taken to extract the remnant mass, because in that regime higher orders in $(\partial_0\phi)^2$ are not negligible anymore.

An important remark concerns the conformal mapping of the CGHS model to another one without kinetic term for the dilaton, a transformation which is frequently used in the literature. Applying the steps above to the latter yields no corrections whatsoever to the Hawking temperature and hence a {\nilcv}. Thus, as might be expected on general grounds (see, for instance, sect.\ 2.1.4 of ref.\ \cite{Grumiller:2002nm}), quantization in different conformal frames yields different results for physical observables.

In previous work most of the calculations of radiative corrections to the Hawking temperature in the CGHS model were done either with some additional assumptions about the quantization (ABC \cite{deAlwis:1992as}) or after a modification of the Polyakov action (BPP \cite{Bose:1995pz,Mikovic:1997de} or RST \cite{Russo:1992ht,Solodukhin:1996te}). 
All of them seem to favor a {\nilcv}. 
These models are exactly soluble\footnote{A very general class
of soluble dilaton models including quantum effects was analyzed
in ref.\ \cite{Zaslavsky:1998ca}.}, 
but their relations to the original one are
somewhat obscure. A direct calculation of the backreaction effects
in the CGHS model with massive scalars was performed in ref.\ \cite{Chiou-Lahanas:1996ea}.
These authors, however, invoked a large mass expansion for the scalars 
which makes any comparison to our calculations hardly possible.

Quantum corrections to mass and temperature have been obtained also in ref.\ \cite{Zaslavskii:1996dg} by thermodynamical methods under the assumption of compactly supported Hawking radiation. In the infinite volume limit the result for the quantum corrected Hawking temperature coincides with (\ref{1loop102}), apart from the sign of the correction term. This sign is very important for physical reasons as it correlates with the sign in (\ref{cghs:1}). While we have obtained a positive specific heat, eq.\ (18) of ref.\ \cite{Zaslavskii:1996dg} implies a negative one. 
%Thus, our calculations indicate that quantum effects stabilize the Hawking flux while Zaslavskii's calculations imply a destabilization. A probable reason of this discrepancy is the fact that there are two contributions to the correction term in (\ref{1loop102}): one arises due to the shift of the horizon, $\exp{(-2\la r_h)}M/\la=1+\la/(48\pi M)$, while the other one originates from the additional contribution to $d\xi/dr=(d\xi/dr)_{0}(1-\exp{(-2\la r)}/(24\pi))$ due to backreaction effects (the subscript ``0'' means leading order). Thus, the total correction is of the form $A(1-2)=-A$. If only the first contribution is considered (as it seems to be the case in ref.\ \cite{Zaslavskii:1996dg}) the sign of the specific heat changes.
The difference may be traced back to the Hartle-Hawking boundary conditions imposed in ref.\ \cite{Zaslavskii:1996dg} which should be supported by the thermal bath, while our discussion is based upon Unruh boundary conditions which seem to be the most adequate ones for describing the evaporation of an isolated macroscopic BH (cf.\ e.g.\ sect.\ 11.2 of ref.\ \cite{Frolov:1998}).

If the CGHS is considered from a stringy 
point of view one should compare it with the exact string BH of ref.\ \cite{Dijkgraaf:1992ba} which follows from an exact conformal field theory. It produces the CGHS in the limit of level $k\to\infty$. But even for arbitrary values of 
$k$ the specific heat does not receive corrections. It should be noted that
the results on the ADM mass and some other characteristics of the exact
string BH should be considered with a certain reservation because no field
theory action for this model is known \cite{Grumiller:2002md}.

Finally, it should be mentioned that the approach advocated here can be applied (at least in principle) to all dilaton gravity theories in {\twod} with minimally coupled matter, albeit complications arise due to the fact that the generalized Polyakov action (\ref{cghs:16}) contains non-local contributions. Moreover, additional diagrams already contribute at tree level \cite{Grumiller:2002dm}. A generalization to nonminimally coupled matter is even less straightforward, but it would be very interesting to pursue, because in this way the genuine Schwarzschild BH with (in four dimensions) minimally coupled matter can be described. 

\section*{Acknowledgement}
 
This work has been supported by project P-14650-TPH of the Austrian Science Foundation (FWF) and by the DFG project 1112/12-1. We are grateful to the referee for his constructive criticism which led to an improvement of the manuscript. We thank O.\ Zaslavskii for bringing ref.\ \cite{Zaslavskii:1996dg} to our attention and for clarifying discussions on its content in numerous e-mails.

\begin{appendix}

\section{Alternative approach}

The purpose of this appendix is to provide an alternative derivation of the effective interaction vertex (\ref{cghs:18}) by applying a method which works also for generic dilaton gravity theories and which has proven useful already for tree level vertices \cite{Kummer:1998zs,Grumiller:2000ah,Grumiller:2001ea,Fischer:2001vz,Grumiller:2002dm}. Since the Polyakov action contains the zweibeine at classical level it is sufficient to solve the classical equations of motion with matter replaced by a localized source $(\partial_0\phi)^2\to c_0\de(x-y)$.
The linear order terms in $c_0$ will produce the vertex of interest. In this manner one obtains\footnote{There are different prescriptions to define $\partial_0^{-2}\de(x-y)$. The one with $\theta(x^0-y^0)$ in (\ref{1loop13}) has been chosen because it produces a finite result for the vertex and thus allows for a sensible discussion of divergencies which are encountered later.} by expanding the logarithm of (\ref{B12})
\eq{
\ln{\hat{E}_1^+}=\ln{E_1^+}-c_0\left(1-\frac{y^0}{x^0}\right)\theta(x^0-y^0)\de(x^1-y^1)+\mathcal{O}(c_0^2)\,.
}{1loop13}
The differential operator $\Ga$ does not receive any $c_0$ corrections. 

As pointed out below eq.\ (\ref{cghs:17}) there seem to be two ways to obtain the vertex: naively, one would just take the first order in $c_0$ of the whole Polyakov action (``symmetric variant''); alternatively, by taking the origin of the Polyakov action, namely the conformal anomaly, seriously one has to take the first order term in $c_0$ of $\ln{\hat{E}_1^+}$ and to multiply it with the zeroth order of the curvature term, $\Ga\ln{\hat{E}_1^+}$ (``correct variant''). The result comprising both cases is
\eq{
V=\int\limits_{-\infty}^{\infty} dy^1\int\limits_{x^0_h+\eps}^{\infty} dy^0 (\partial_0 \phi)^2 \left[\frac{c}{y^0}+d\right]\,.
}{1loop15}
The symmetric variant yields $c=-M/(192\pi\la^2)$ and $d=1/48\pi$ while otherwise $c=-M/(192\pi\la^2)$ and $d=0$ is obtained. The lower integration limit is explained as follows: $x^0_h=M/\la$ is the horizon of the background geometry and $\eps>0$ is a cutoff parameter to regularize the $y^0$ integration in (\ref{1loop15}).

For $d=0$ the result (\ref{1loop15}) is equivalent to (\ref{cghs:18}). To show that $d$ must be zero it is sufficient to study propagation in the vacuum, i.e.\ a vanishing BH mass $M=0$ can be considered. Then $c=0$, but $d$ might be vanishing or nonvanishing, depending on whether the ``correct'' or the ``symmetric'' variant is applied. The scalar field can be represented as
\meq{
\phi=\frac{1}{\sqrt{2\pi}}\int\limits_0^\infty \frac{dk}{\sqrt{2k}} \Bigg[b^+_k\exp{(ikx^1)}+b^-_k\exp{(-ikx^1)} \\
+a_k^+\exp{\left(ik\left(x^1+\int^{x^0}\frac{dz}{\la z}\right)\right)}+a_k^-\exp{\left(-ik\left(x^1+\int^{x^0}\frac{dz}{\la z}\right)\right)}\Bigg]\,,
}{1loop17}
where $b^\pm$ create and annihilate the right movers and $a^\pm$ the left movers. The normalization is given by $[a_k^-,a_q^+]=\de(k-q)=[b_k^-,b_q^+]$ (all other commutators vanish).

In the quantity $(\partial_0\phi)^2$ only the left movers survive; therefore, only left movers may acquire nontrivial quantum corrections. The relation between in and out vacua is trivial and the T-matrix corresponding to a propagator correction can be calculated straightforwardly:
\eq{
T=_{\rm out}<0|a^-_k V a^+_q|0>_{\rm in} \propto \de(k-q) \frac{d}{\eps}
}{1loop16}
Thus, in the correct variant no additional correction to the propagator arises, while the symmetric variant yields a cutoff dependent result which diverges when the cutoff approaches zero. This provides a physical way to see why the ``correct'' variant can be the only valid one: after all, for vanishing BH mass ($M=0$) the scalar field should propagate freely on a Minkowskian background.

In the massive case ($M\neq 0$) the standard route would be to introduce a mode decomposition for the scalar field (carefully distinguishing between in modes and out modes) and to calculate vacuum expectation values with the insertion of (\ref{cghs:18}). Assuming that $V$ is a small perturbation standard methods can be applied and then the problem reduces to the determination of quantum corrected Bogoliubov coefficients between in states and out states. Once these coefficients are known also corrections to the Hawking temperature can be extracted. However, there are difficulties as the vertex introduced in (\ref{1loop15}) depends on the cutoff parameter $\eps$ and diverges in the limit $\eps\to 0$. Therefore, the method presented in the main part is more suitable if one wants to calculate just corrections to the Hawking temperature rather than generic scattering processes.

\end{appendix}

%%% END OF PAPER %%%

%%% REFERENCES %%%

%\bibliographystyle{../REV/thisisthedirectoryofdima/fullcream} % for the author's convenience
%\bibliography{review,dv} % my bibtex-file - just use the .bbl file instead or I can send you my bibtex file, if you want (but please do not include new refs in it - rather let me include them, it is much easier to handle this ``centralized''

%\bibliographystyle{JHEP} % for the author's convenience
%\bibliography{../review01/review} % my bibtex-file - just use the .bbl file instead or I can send you my bibtex file, if you want (but please do not include new refs in it - rather let me include them, it is much easier to handle this ``centralized''

\begin{thebibliography}{10}

\bibitem{Smolin:2003rk}
L.~Smolin, {\it How far are we from the quantum theory of gravity?},
  \href{http://xxx.lanl.gov/abs/hep-th/0303185}{{\tt hep-th/0303185}}.

\bibitem{Grumiller:2002nm}
D.~Grumiller, W.~Kummer, and D.~V. Vassilevich, {\it Dilaton gravity in two
  dimensions},  {\em Phys. Rept.} {\bf 369} (2002) 327--429,
  [\href{http://xxx.lanl.gov/abs/http://arXiv.org/abs/hep-th/0204253}{{\tt
  http://arXiv.org/abs/hep-th/0204253}}].

\bibitem{Callan:1992rs}
C.~G. Callan, Jr., S.~B. Giddings, J.~A. Harvey, and A.~Strominger, {\it
  Evanescent black holes},  {\em Phys. Rev.} {\bf D45} (1992) 1005--1009,
  [\href{http://xxx.lanl.gov/abs/hep-th/9111056}{{\tt hep-th/9111056}}].

\bibitem{Mandal:1991tz}
G.~Mandal, A.~M. Sengupta, and S.~R. Wadia, {\it Classical solutions of
  two-dimensional string theory},  {\em Mod. Phys. Lett.} {\bf A6} (1991)
  1685--1692;
%\bibitem{Elitzur:1991cb}
S.~Elitzur, A.~Forge, and E.~Rabinovici, {\it Some global aspects of string
  compactifications},  {\em Nucl. Phys.} {\bf B359} (1991) 581--610;
%\bibitem{Witten:1991yr}
E.~Witten, {\it On string theory and black holes},  {\em Phys. Rev.} {\bf D44}
  (1991) 314--324.

\bibitem{Dijkgraaf:1992ba}
R.~Dijkgraaf, H.~Verlinde, and E.~Verlinde, {\it String propagation in a black
  hole geometry},  {\em Nucl. Phys.} {\bf B371} (1992) 269--314.

\bibitem{Giddings:1994pj}
S.~B. Giddings, {\it Quantum mechanics of black holes},  {\em Trieste HEP
  Cosmology} (1994) 0530--574,
  [\href{http://xxx.lanl.gov/abs/arXiv:hep-th/9412138}{{\tt
  arXiv:hep-th/9412138}}];
%\bibitem{Strominger:1994tn}
A.~Strominger, {\it Les {H}ouches lectures on black holes},
  \href{http://xxx.lanl.gov/abs/arXiv:hep-th/9501071}{{\tt
  arXiv:hep-th/9501071}}. Talk given at {NATO} Advanced Study Institute;
%\bibitem{Chamseddine:1992qu}
A.~H. Chamseddine, {\it A study of noncritical strings in arbitrary
  dimensions},  {\em Nucl. Phys.} {\bf B368} (1992) 98--120.

\bibitem{deAlwis:1992as}
S.~P. de~Alwis, {\it Quantization of a theory of 2-d dilaton gravity},  {\em
  Phys. Lett.} {\bf B289} (1992) 278--282,
  [\href{http://xxx.lanl.gov/abs/hep-th/9205069}{{\tt hep-th/9205069}}];
%\bibitem{deAlwis:1993zy}
% S.~P. de~Alwis, 
{\it {Black hole physics from Liouville theory}},  {\em Phys.
  Lett.} {\bf B300} (1993) 330--335,
  [\href{http://xxx.lanl.gov/abs/http://arXiv.org/abs/hep-th/9206020}{{\tt
  http://arXiv.org/abs/hep-th/9206020}}];
%\bibitem{Bilal:1993kv}
A.~Bilal and C.~G. Callan, {\it Liouville models of black hole evaporation},
  {\em Nucl. Phys.} {\bf B394} (1993) 73--100,
  [\href{http://xxx.lanl.gov/abs/http://arXiv.org/abs/hep-th/9205089}{{\tt
  http://arXiv.org/abs/hep-th/9205089}}].

\bibitem{Bose:1995pz}
S.~Bose, L.~Parker, and Y.~Peleg, {\it Semiinfinite throat as the end state
  geometry of two- dimensional black hole evaporation},  {\em Phys. Rev.} {\bf
  D52} (1995) 3512--3517,
  [\href{http://xxx.lanl.gov/abs/http://arXiv.org/abs/hep-th/9502098}{{\tt
  http://arXiv.org/abs/hep-th/9502098}}].

\bibitem{Russo:1992ht}
J.~G. Russo, L.~Susskind, and L.~Thorlacius, {\it Black hole evaporation in
  $1+1$ dimensions},  {\em Phys. Lett.} {\bf B292} (1992) 13--18,
  [\href{http://xxx.lanl.gov/abs/hep-th/9201074}{{\tt hep-th/9201074}}];
%\bibitem{Russo:1992ax}
% J.~G. Russo, L.~Susskind, and L.~Thorlacius, 
{\it {The Endpoint of Hawking
  radiation}},  {\em Phys. Rev.} {\bf D46} (1992) 3444--3449,
  [\href{http://xxx.lanl.gov/abs/http://arXiv.org/abs/hep-th/9206070}{{\tt
  http://arXiv.org/abs/hep-th/9206070}}];
%\bibitem{Russo:1993yh}
% J.~G. Russo, L.~Susskind, and L.~Thorlacius, 
{\it Cosmic censorship in
  two-dimensional gravity},  {\em Phys. Rev.} {\bf D47} (1993) 533--539,
  [\href{http://xxx.lanl.gov/abs/hep-th/9209012}{{\tt hep-th/9209012}}].

\bibitem{Mikovic:1992id}
A.~Mikovic, {\it Exactly solvable models of 2-d dilaton quantum gravity},  {\em
  Phys. Lett.} {\bf B291} (1992) 19--25,
  [\href{http://xxx.lanl.gov/abs/hep-th/9207006}{{\tt hep-th/9207006}}];
%\bibitem{Cangemi:1996yz}
D.~Cangemi, R.~Jackiw, and B.~Zwiebach, {\it Physical states in matter coupled
  dilaton gravity},  {\em Ann. Phys.} {\bf 245} (1996) 408--444,
  [\href{http://xxx.lanl.gov/abs/hep-th/9505161}{{\tt hep-th/9505161}}];
%\bibitem{Benedict:1996qy}
E.~Benedict, R.~Jackiw, and H.~J. Lee, {\it Functional {S}chroedinger and
  {BRST} quantization of (1+1)- dimensional gravity},  {\em Phys. Rev.} {\bf
  D54} (1996) 6213--6225, [\href{http://xxx.lanl.gov/abs/hep-th/9607062}{{\tt
  hep-th/9607062}}].

\bibitem{Mikovic:1997de}
A.~Mikovic and V.~Radovanovic, {\it Loop corrections in the spectrum of 2d
  {H}awking radiation},  {\em Class. Quant. Grav.} {\bf 14} (1997) 2647--2661,
  [\href{http://xxx.lanl.gov/abs/gr-qc/9703035}{{\tt gr-qc/9703035}}].

\bibitem{Kuchar:1997zm}
K.~V. Kucha{\v{r}}, J.~D. Romano, and M.~Varadarajan, {\it Dirac constraint
  quantization of a dilatonic model of gravitational collapse},  {\em Phys.
  Rev.} {\bf D55} (1997) 795--808,
  [\href{http://xxx.lanl.gov/abs/gr-qc/9608011}{{\tt gr-qc/9608011}}];
%\bibitem{Varadarajan:1998qz}
M.~Varadarajan, {\it Quantum gravity effects in the {CGHS} model of collapse to
  a black hole},  {\em Phys. Rev.} {\bf D57} (1998) 3463--3473,
  [\href{http://xxx.lanl.gov/abs/gr-qc/9801058}{{\tt gr-qc/9801058}}].

\bibitem{Haider:1994cw}
F.~Haider and W.~Kummer, {\it {Quantum functional integration of nonEinsteinian
  gravity in d = 2}},  {\em Int. J. Mod. Phys.} {\bf A9} (1994) 207--220.

\bibitem{Kummer:1997hy}
W.~Kummer, H.~Liebl, and D.~V. Vassilevich, {\it Exact path integral
  quantization of generic 2-d dilaton gravity},  {\em Nucl. Phys.} {\bf B493}
  (1997) 491--502,
  [\href{http://xxx.lanl.gov/abs/http://arXiv.org/abs/gr-qc/9612012}{{\tt
  http://arXiv.org/abs/gr-qc/9612012}}].

\bibitem{Kummer:1998zs}
W.~Kummer, H.~Liebl, and D.~V. Vassilevich, {\it Integrating geometry in
  general 2d dilaton gravity with matter},  {\em Nucl. Phys.} {\bf B544} (1999)
  403--431, [\href{http://xxx.lanl.gov/abs/hep-th/9809168}{{\tt
  hep-th/9809168}}].

\bibitem{Grumiller:2000ah}
D.~Grumiller, W.~Kummer, and D.~V. Vassilevich, {\it The virtual black hole in
  2d quantum gravity},  {\em Nucl. Phys.} {\bf B580} (2000) 438--456,
  [\href{http://xxx.lanl.gov/abs/gr-qc/0001038}{{\tt gr-qc/0001038}}].

\bibitem{Grumiller:2001ea}
D.~Grumiller, {\em Quantum dilaton gravity in two dimensions with matter}.
\newblock PhD thesis, {T}echnische {U}niversit{\"a}t {W}ien, 2001.
\newblock \href{http://xxx.lanl.gov/abs/gr-qc/0105078}{{\tt gr-qc/0105078}}.

\bibitem{Frolov:1998}
V.~Frolov and I.~Novikov, {\em {Black Hole Physics}}.
\newblock Kluwer Academic Publishers, 1998.

\bibitem{Katanaev:1996bh}
M.~O. Katanaev, W.~Kummer, and H.~Liebl, {\it {Geometric Interpretation and
  Classification of Global Solutions in Generalized Dilaton Gravity}},  {\em
  Phys. Rev.} {\bf D53} (1996) 5609--5618,
  [\href{http://xxx.lanl.gov/abs/http://arXiv.org/abs/gr-qc/9511009}{{\tt
  http://arXiv.org/abs/gr-qc/9511009}}];
%\bibitem{Katanaev:1997ni}
% M.~O. Katanaev, W.~Kummer, and H.~Liebl, 
{\it On the completeness of the black
  hole singularity in 2d dilaton theories},  {\em Nucl. Phys.} {\bf B486}
  (1997) 353--370, [\href{http://xxx.lanl.gov/abs/gr-qc/9602040}{{\tt
  gr-qc/9602040}}].

\bibitem{Ikeda:1993aj}
N.~Ikeda and K.~I. Izawa, {\it General form of dilaton gravity and nonlinear
  gauge theory},  {\em Prog. Theor. Phys.} {\bf 90} (1993) 237--246,
  [\href{http://xxx.lanl.gov/abs/hep-th/9304012}{{\tt hep-th/9304012}}];
%\bibitem{Ikeda:1994fh}
N.~Ikeda, {\it Two-dimensional gravity and nonlinear gauge theory},  {\em Ann.
  Phys.} {\bf 235} (1994) 435--464,
  [\href{http://xxx.lanl.gov/abs/arXiv:hep-th/9312059}{{\tt
  arXiv:hep-th/9312059}}];
%\bibitem{Schaller:1994es}
P.~Schaller and T.~Strobl, {\it Poisson structure induced (topological) field
  theories},  {\em Mod. Phys. Lett.} {\bf A9} (1994) 3129--3136,
  [\href{http://xxx.lanl.gov/abs/http://arXiv.org/abs/hep-th/9405110}{{\tt
  http://arXiv.org/abs/hep-th/9405110}}].

\bibitem{Klosch:1996fi}
T.~Kl{\"o}sch and T.~Strobl, {\it Classical and quantum gravity in
  (1+1)-dimensions. {P}art 1: {A} unifying approach},  {\em Class. Quant.
  Grav.} {\bf 13} (1996) 965--984,
  [\href{http://xxx.lanl.gov/abs/arXiv:gr-qc/9508020}{{\tt
  arXiv:gr-qc/9508020}}];
%\bibitem{Klosch:1996qv}
% T.~Kl{\"o}sch and T.~Strobl, 
{\it Classical and quantum gravity in 1+1
  dimensions. Part {II}: {T}he universal coverings},  {\em Class. Quant. Grav.}
  {\bf 13} (1996) 2395--2422,
  [\href{http://xxx.lanl.gov/abs/arXiv:gr-qc/9511081}{{\tt
  arXiv:gr-qc/9511081}}];
%\bibitem{Klosch:1997md}
% T.~Kl{\"o}sch and T.~Strobl, 
{\it {Classical and quantum gravity in 1+1
  dimensions. Part III: Solutions of arbitrary topology and kinks in 1+1
  gravity}},  {\em Class. Quant. Grav.} {\bf 14} (1997) 1689--1723,
  [\href{http://xxx.lanl.gov/abs/http://arXiv.org/abs/hep-th/9607226}{{\tt
  http://arXiv.org/abs/hep-th/9607226}}].

\bibitem{Fradkin:1975cq}
E.~S. Fradkin and G.~A. Vilkovisky, {\it Quantization of relativistic systems
  with constraints},  {\em Phys. Lett.} {\bf B55} (1975) 224;
%\bibitem{Batalin:1977pb}
I.~A. Batalin and G.~A. Vilkovisky, {\it Relativistic {S} matrix of dynamical
  systems with boson and fermion constraints},  {\em Phys. Lett.} {\bf B69}
  (1977) 309--312;
%\bibitem{Fradkin:1978xi}
E.~S. Fradkin and T.~E. Fradkina, {\it Quantization of relativistic systems
  with boson and fermion first and second class constraints},  {\em Phys.
  Lett.} {\bf B72} (1978) 343.

\bibitem{Fujikawa:1988ie}
K.~Fujikawa, U.~Lindstrom, N.~K. Nielsen, M.~Rocek, and P.~van Nieuwenhuizen,
  {\it The regularized {BRST} coordinate invariant measure},  {\em Phys. Rev.}
  {\bf D37} (1988) 391.

\bibitem{Katanaev:2000kc}
M.~O. Katanaev, {\it Effective action for scalar fields in two-dimensional
  gravity},  {\em Annals Phys.} {\bf 296} (2002) 1--50,
  [\href{http://xxx.lanl.gov/abs/http://arXiv.org/abs/gr-qc/0101033}{{\tt
  http://arXiv.org/abs/gr-qc/0101033}}].

\bibitem{Fischer:2001vz}
P.~Fischer, D.~Grumiller, W.~Kummer, and D.~V. Vassilevich, {\it S-matrix for
  s-wave gravitational scattering},  {\em Phys. Lett.} {\bf B521} (2001)
  357--363,
  [\href{http://xxx.lanl.gov/abs/http://arXiv.org/abs/gr-qc/0105034}{{\tt
  http://arXiv.org/abs/gr-qc/0105034}}]. Erratum ibid. {\bf B532} (2002) 373;
%\bibitem{Grumiller:2001rg}
D.~Grumiller, {\it Virtual black hole phenomenology from 2d dilaton theories},
  {\em Class. Quant. Grav.} {\bf 19} (2002) 997--1009,
  [\href{http://xxx.lanl.gov/abs/http://arXiv.org/abs/gr-qc/0111097}{{\tt
  http://arXiv.org/abs/gr-qc/0111097}}].

\bibitem{Grumiller:2002dm}
D.~Grumiller, W.~Kummer, and D.~V. Vassilevich, {\it Virtual black holes in
  generalized dilaton theories and their special role in string gravity},
  \href{http://xxx.lanl.gov/abs/http://arXiv.org/abs/hep-th/0208052}{{\tt
  http://arXiv.org/abs/hep-th/0208052}}.

\bibitem{Polyakov:1981rd}
A.~M. Polyakov, {\it Quantum geometry of bosonic strings},  {\em Phys. Lett.}
  {\bf B103} (1981) 207--210.

\bibitem{Dowker:1994rt}
J.~S. Dowker, {\it A note on {P}olyakov's nonlocal form of the effective
  action},  {\em Class. Quant. Grav.} {\bf 11} (1994) L7--L10,
  [\href{http://xxx.lanl.gov/abs/hep-th/9309127}{{\tt hep-th/9309127}}].

\bibitem{Kummer:1999zy}
W.~Kummer and D.~V. Vassilevich, {\it {Hawking radiation from dilaton gravity
  in (1+1) dimensions: A pedagogical review}},  {\em Annalen Phys.} {\bf 8}
  (1999) 801--827,
  [\href{http://arXiv.org/abs/gr-qc/9907041}{{\tt
  http://arXiv.org/abs/gr-qc/9907041}}].

\bibitem{Grumiller:1999rz}
D.~Grumiller and W.~Kummer, {\it Absolute conservation law for black holes},
{\em Phys.\ Rev.} {\bf D61} (2000) 064006,
[\href{http://arXiv.org/abs/gr-qc/9902074}{{\tt http://arXiv.org/abs/gr-qc/9902074}}].
%%CITATION = GR-QC 9902074;%%

\bibitem{Corless:1996}
R.~M. Corless, G.~H. Gonnet, D.~E.~G. Hare, D.~J. Jeffrey, and D.~E. Knuth,
  {\it On the Lambert W Function}, {\em Adv. Comp. Math.} {\bf 5} (1996)
  329--359. More information on the Lambert $W$ function is available at the
  webpage
  \href{http://kong.apmaths.uwo.ca/~rcorless/frames/PAPERS/LambertW/}{{\tt
  http://kong.apmaths.uwo.ca/$\sim$rcorless/frames/PAPERS/LambertW/}}.

\bibitem{Solodukhin:1996te}
S.~N. Solodukhin, {\it Two-dimensional quantum corrected eternal black hole},
  {\em Phys. Rev.} {\bf D53} (1996) 824--835,
  [\href{http://xxx.lanl.gov/abs/hep-th/9506206}{{\tt hep-th/9506206}}].

\bibitem{Zaslavsky:1998ca}
O.~B. Zaslavsky, {\it Exactly solvable models of two-dimensional dilaton
  gravity and quantum eternal black holes},  {\em Phys. Rev.} {\bf D59} (1999)
  084013,
  [\href{http://xxx.lanl.gov/abs/http://arXiv.org/abs/hep-th/9804089}{{\tt
  http://arXiv.org/abs/hep-th/9804089}}].

\bibitem{Chiou-Lahanas:1996ea}
C.~Chiou-Lahanas, G.~A. Diamandis, B.~C. Georgalas, A.~Kapella-Economou, and
  X.~N. Maintas, {\it Backreaction effect in the two-dimensional dilaton
  gravity},  {\em Phys. Rev.} {\bf D54} (1996) 6226--6232,
  [\href{http://xxx.lanl.gov/abs/hep-th/9607117}{{\tt hep-th/9607117}}].

\bibitem{Zaslavskii:1996dg}
O.~B. Zaslavskii, {\it Quantum corrections to temperature and mass of 1+1 dilatonic black holes and the trace anomaly}, {\em Phys. Lett.} {\bf B375} (1996) 43--46.

\bibitem{Grumiller:2002md}
D.~Grumiller and D.~V. Vassilevich, {\it Non-existence of a dilaton gravity
  action for the exact string black hole},  {\em JHEP} {\bf 11} (2002) 018,
  [\href{http://xxx.lanl.gov/abs/http://arXiv.org/abs/hep-th/0210060}{{\tt
  http://arXiv.org/abs/hep-th/0210060}}].

\end{thebibliography}

\providecommand{\href}[2]{#2}\begingroup\raggedright\endgroup

\end{document}